\documentclass[aps,prl,showpacs,twocolumn,floats]{revtex4}
\usepackage{amssymb}

\usepackage{graphicx}
\usepackage{dcolumn}
\usepackage{bm}

\begin{document}

\bibliographystyle{prsty}
\author{Reem Jaafar and E. M. Chudnovsky}
\affiliation{Physics Department, Lehman College, The City
University of New York \\ 250 Bedford Park Boulevard West, Bronx,
New York 10468-1589, U.S.A.}
\date{\today}

\begin{abstract}
We study the quantum dynamics of a system consisting of a magnetic
molecule placed on a microcantilever. The amplitude and
frequencies of the coupled magneto-mechanical oscillations are
computed. Parameter-free theory shows that the existing
experimental techniques permit observation of the driven coupled
oscillations of the spin and the cantilever, as well as of the
splitting of the mechanical modes of the cantilever caused by spin
tunneling.
\end{abstract}

\title{Magnetic Molecule on a Microcantilever: Quantum Magneto-Mechanical Oscillations}

\maketitle

Magnetic molecules exhibit quantum tunneling between different
orientations of the spin in macroscopic magnetization measurements
\cite{Friedman,book1}. Detection of coherent quantum spin
oscillations, similar to those observed in a SQUID \cite{SQUID},
would be of great interest. In a crystal of magnetic molecules
this effect is difficult to observe because of the inhomogeneous
broadening of spin levels and decoherence arising from various
interactions. Magnetic measurements of individual molecules would
have been more promising but insufficient sensitivity of existing
magnetometers has prohibited such studies so far. The effort has
been made to observe spin tunneling effects in the electron
transport through a single magnetic molecule bridged between
metallic electrodes \cite{transport}.

\begin{figure}[ht]
\begin{center}
\vspace{-0.5cm}
\includegraphics[width=67.8mm,angle=-90]{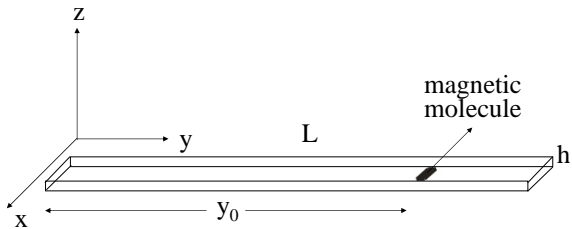}
\vspace{-3cm} \caption{Geometry considered in the paper.}
\label{geometry}
\end{center}
\end{figure}

In this Letter we propose a different approach to the detection of
quantum oscillations of the spin of a single magnetic molecule. It
is based upon the resonant coupling of the spin oscillations to
the mechanical modes of a microcantilever. The geometry of the
proposed experiment is shown in Fig.\ \ref{geometry}.
A magnetic molecule of spin ${\bf S}$ is deposited on a
microcantilever of length $L$, Fig. 1. The $y = 0$ end of the
cantilever is fixed while the $y = L$ end is free. The molecule is
assumed to be imbedded in or firmly attached to the cantilever at
$y = y_0$, with the magnetic anisotropy axis being parallel the
$X$-direction. Let $\Delta$ be the tunnel splitting of the two
lowest energy states of the molecule,
\begin{equation}\label{Psi}
|\Psi_{\pm}\rangle = \frac{1}{\sqrt{2}}\left(|S\rangle \pm
|-S\rangle\right)\,,
\end{equation}
where $|\pm S\rangle$ denote two opposite spin orientations along
the $X$-axis. A weak ac magnetic field of frequency $\omega =
\Delta/\hbar$, applied along the $X$-axis, will force the spin of
the molecule to oscillate between the two orientations.
Conservation of the total angular momentum requires that the
oscillations of the spin are accompanied by the mechanical
oscillations of the cantilever (Einstein - de Haas effect
\cite{EdH,NIST}). Consequently, if $\omega = \Delta/\hbar$
coincides with a resonant mode of the cantilever, one should
expect the effect of the ac field on the cantilever.

If the cantilever rotates by a small angle $\delta \bm{\phi}$ at
the location of the spin, the Hamiltonian of the molecule
$\hat{H}_{S}$ becomes \cite{Fulde,CGS}
\begin{equation}\label{R-A}
\hat{H}'_{S}= \hat{R}\hat{H}_{S}\hat{R}^{-1} =
\hat{H}_{S}\mathbf{+}i\sum\left[ \hat{H}_{S},\mathbf{S} \right]
\cdot \delta \bm{\phi} = \hbar \dot{\bf S} \cdot \delta {\bm
{\phi}}\,. \label{Smallphi}
\end{equation}
Due to strong magnetic anisotropy the molecule can be considered
as a two-state system. The Hamiltonian of such a system,
${\cal{H}}_2$, is a projection of the Hamiltonian (\ref{R-A}) onto
the two states given by Eq.\ (\ref{Psi}). These states can be
viewed as the eigenstates of the Pauli matrix $\sigma_z$. For the
geometry shown in Fig.\ \ref{geometry} the projection can be
performed by writing
\begin{equation}
{\cal{H}}_2 = \sum_{i,j=\pm}\langle \Psi_i|\left(\hat{\cal{H}}_S+
i\left[\hat{\cal{H}}_S,S_x\right]\delta\phi_x\right)|\Psi_j\rangle|\Psi_i\rangle\langle\Psi_j|
\end{equation}
with
\begin{eqnarray}
\sigma_x & = &
|\Psi_+\rangle\langle\Psi_-|+|\Psi_-\rangle\langle\Psi_+| \label{sigmax-psi}\\
\sigma_y & = &-i|\Psi_+\rangle\langle\Psi_-|+i|\Psi_-\rangle\langle\Psi_+|\label{sigmay-psi}\\
\sigma_z & = & |\Psi_+\rangle\langle\Psi_+| -
|\Psi_-\rangle\langle\Psi_-|\label{sigmaz-psi}\,.
\end{eqnarray}
This gives
\begin{equation}
{\cal{H}}_2 = -\frac{1}{2} \Delta \sigma_z + S \Delta \delta\phi_x
\sigma_y\,.
\end{equation}
The states with a definite $X$-projection of the spin are
eigenstates of $\sigma_x$:  $| \pm S\rangle =
\frac{1}{\sqrt{2}}(|\Psi_+\rangle \pm |\Psi_-\rangle)$. The
expectation value of ${\bm \sigma}$ satisfies the Landau-Lifshitz
equation:
\begin{equation}\label{LL-sigma}
\hbar \dot{\bm \sigma} = -{\bm \sigma} \times {\bf b}_{\rm eff}\,,
\end{equation}
with
\begin{equation}\label{b-eff}
{\bf b}_{\rm eff}  = -\Delta {\bf e}_z + 2 S \Delta \delta \phi_x
{\bf e}_y\,.
\end{equation}
If the cantilever was held stationary ($\delta \phi_x=0$), the
solution of Eq.\ (\ref{LL-sigma}) would describe pure quantum
oscillations of the spin: $\sigma_z =  const, \sigma_x \propto
\cos({\Delta}{t}/\hbar), \sigma_y \propto
\sin({\Delta}{t}/\hbar)$.

We are interested in the coupled oscillations of the spin and the
cantilever. Since the latter contains macroscopic number of atoms,
its oscillations can be studied within a continuous elastic theory
that deals with the displacement field ${\bf u}({\bf r},t)$ and
the local rotation $\delta \bm{\phi}$ given by \cite{LL}
\begin{equation}\label{phi-u}
\delta \mathbf{\phi (\mathbf{r})=}\frac{1}{2}\nabla \times \mathbf{u}(%
\mathbf{r})  \,.
\end{equation}
Within such a model the spin of the molecule at a point ${\bf
r}_0$ can be replaced by the spin field
\begin{equation}
{\bf \Sigma}({\bf r},t) = S{\bm \sigma}(t)\delta ({\bf r} - {\bf
r}_0)\,,
\end{equation}
where ${\bm \sigma}(t)$ satisfies Eq.\ (\ref{LL-sigma}). With
account of Eq.\ (\ref{R-A}) the energy of the cantilever becomes
\begin{equation}\label{H}
{\cal{H}}_C = {\cal{H}}_{E} + \frac{1}{2}\int d^3r \,\hbar
\dot{\bf \Sigma}  \cdot (\nabla \times \bf{u})\,,
\end{equation}
where ${\cal{H}}_{E}$ is the part of the elastic energy that is
independent of the spin.

The dynamical equation for the displacement field is
\begin{equation}\label{elastic}
\rho \frac{\partial^2 u_{\alpha}}{\partial t^2} = \frac{\partial
\sigma_{\alpha \beta}}{\partial x_{\beta}} \,,
\end{equation}
where $\sigma_{\alpha \beta} = {\delta {\cal{H}}_C}/\delta
e_{\alpha \beta}$ is the stress tensor, $e_{\alpha \beta} =
\partial u_{\alpha}/\partial x_{\beta}$ is the strain tensor, and
$\rho$ is the mass density of the material. This gives
\begin{equation}\label{torque}
\rho \frac{\partial^2 u_{\alpha}}{\partial t^2} - \frac{\partial
\sigma^{(\rm{E})}_{\alpha \beta}}{\partial x_{\beta}} = -
\frac{\hbar}{2}{\bm \nabla}\times \dot{\bf \Sigma}
\end{equation}
where $\sigma^{(\rm{E})}_{\alpha \beta} = {\delta {\cal{H}}_{E}
}/{\delta e_{\alpha \beta}}$. It is easy to see that in the
absence of the external torque, ${K}_{\alpha}^{(E)} = \oint d
A_{\delta}\left[\epsilon_{\alpha\beta\gamma}r_{\beta}\sigma^{(E)}_{\gamma\delta}\right]$,
applied to the surface of the body ${\bf A}$,  Eq.\ (\ref{torque})
provides conservation of the total angular momentum,
\begin{equation}\label{J}
{\bf J} = \int d^3r \left[ \hbar{\bf \Sigma} + \rho\left({\bf r}
\times \dot{\bf u}\right)\right]\,,\quad d{\bf J}/dt = 0\,.
\end{equation}

Writing the left-hand side of Eq.\ (\ref{torque}) in the
conventional form for a cantilever \cite{LL} one obtains the
elastic equation that couples vertical displacements of the
cantilever, $u_z(y,t)$, with the oscillations of the spin:
\begin{equation}\label{elastic-S}
\rho \frac{\partial^2 u_z}{\partial t^2} +
\frac{h^2E}{12(1-\sigma^2)}\frac{\partial^4 u_z}{\partial y^4} =
\frac{\hbar S}{2V}\frac{\partial}{\partial y}\frac{\partial
}{\partial t}\left[\sigma_{x}(t)L\delta(y - y_0)\right]\,.
\end{equation}
Here $h$ and $V$ are the thickness and the volume of the
cantilever, respectively, $E$ is the Young's modulus, and $\sigma$
is the Poisson coefficient, $-1 < \sigma < 1/2$.

It is convenient to switch in Eq.\ (\ref{elastic-S}) to
dimensionless variables $\bar{u}_z = {u_z}/{L}, \bar{y} = {y}/{L},
\bar{t} = t\nu$, where
\begin{equation}\label{bar}
\nu \equiv \sqrt{\frac{Eh^2}{12\rho(1-\sigma^2)L^4}}
\end{equation}
determines the scale of the eigenfrequencies of the oscillations
of the cantilever. By order of magnitude $\nu \sim v_sh/L^2$ where
$v_s \sim \sqrt{E/\rho}$ is the speed of sound. In terms of these
variables Eq.\ (\ref{elastic-S}) becomes
\begin{equation}\label{eq-bar}
\frac{\partial^2 \bar{u}_z}{\partial \bar{t}^2} + \frac{\partial^4
\bar{u}_z}{\partial \bar{y}^4} = \frac{\epsilon}{2}
\frac{\partial}{\partial \bar{y}}\frac{\partial }{\partial
\bar{t}}\left[\sigma_{x}(\bar{t})\delta(\bar{y} -
\bar{y}_0)\right]\,,
\end{equation}
where $0< \bar{y}_0 <1$ and
\begin{equation}\label{epsilon}
\epsilon = \frac{\hbar S}{M L^2\nu}= \frac{\hbar
S}{M}\sqrt{\frac{12\rho(1-\sigma^2)}{Eh^2}}
\end{equation}
is a dimensionless small parameter. By order of magnitude,
$\epsilon \sim \hbar S /(Mv_sh)$, where $M = \rho V$ is the mass
of the cantilever. For, e.g., a molecule of spin $S=10$ on a
cantilever of dimensions 100nm$\times$10nm$\times$1nm the
parameter $\epsilon$ should be of order $10^{-7}$. Eq.\
(\ref{eq-bar}) has to be solved with the following boundary
conditions:
\begin{eqnarray}\label{boundary}
& & \bar{u}_z = 0\,,\; \frac{\partial \bar{u}_z}{\partial \bar{y}}
= 0\;\;\; {\rm at}\;\;\; \bar{y} = 0\,, \nonumber \\
& &
\frac{\partial^2 \bar{u}_z}{\partial \bar{y}^2} = 0\,,\;
\frac{\partial^3 \bar{u}_z}{\partial \bar{y}^3} = 0\;\;\; {\rm
at}\;\;\; \bar{y} = 1\,.
\end{eqnarray}
The first two conditions correspond to the absence of displacement
and the absence of bending of the cantilever at the fixed end,
while the last two conditions correspond to the absence of torque
and force, respectively, at the free end \cite{LL}.

\begin{figure}[ht]
\begin{center}
\includegraphics[width=65mm, angle = -90]{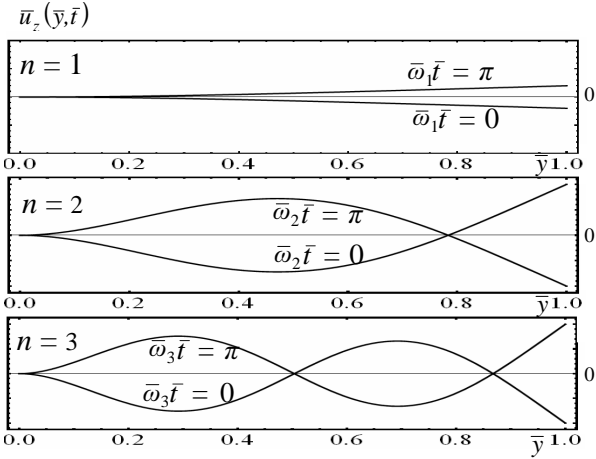}
\caption{Profiles of the oscillating cantilever at different
moments of time for $n = 1,2,3$. } \label{Free}
\end{center}
\end{figure}

For the free oscillations of the cantilever ($\epsilon = 0$) one
writes
\begin{equation}\label{free}
\bar{u}_z(\bar{y}, \bar{t}) =
\bar{u}(\bar{y})\cos(\bar{\omega}\bar{t})\,.
\end{equation}
Substitution into Eq.\ (\ref{eq-bar}) with $\epsilon = 0$ then
gives
\begin{equation}\label{eq-u}
\frac{\partial^4 \bar{u}}{\partial \bar{y}^4} - \kappa^4\bar{u} =
0\,, \qquad \kappa^2 \equiv \bar{\omega}\,.
\end{equation}
Solutions are
\begin{eqnarray}\label{eigenfunction}
& & \bar{u}(\bar{y}) = (\cos\kappa + \cosh\kappa)\left[\cos(\kappa
\bar{y}) - \cosh(\kappa
\bar{y})\right] \nonumber \\
& & + (\sin\kappa - \sinh\kappa)\left[\sin(\kappa \bar{y})-
\sinh(\kappa \bar{y})\right]\,.
\end{eqnarray}
The third of the boundary conditions (\ref{boundary}) provides the
equation,
\begin{equation}\label{modes}
\cos\kappa\cosh\kappa + 1 = 0\,,
\end{equation}
for the frequencies of the normal modes of the cantilever,
$\bar{\omega}_n = \kappa^2_n$ (measured in the units of $\nu$ of
Eq.\ (\ref{bar})). The fundamental (minimal) frequency is
$\bar{\omega}_1 \approx 3.516$. The next two frequencies are
$\bar{\omega}_2 \approx 22.03$ and $\bar{\omega}_3 \approx 61.70$.
The profiles of the oscillations of the cantilever for three
normal modes ($n=1,2,3$) are shown in Fig. \ref{Free}.

To consider coupled oscillations of the cantilever and the spin of
the molecule we first neglect dissipation and write for the
displacement
\begin{equation}\label{forced}
\bar{u}_z(\bar{y}, \bar{t}) =
\sum_mR_m(\bar{t})\bar{u}_m(\bar{y})\,,
\end{equation}
where $R_m(t)$ are functions of time to be determined and
$\bar{u}_m(\bar{y})$ is a normalized eigenfunction
(\ref{eigenfunction}) that corresponds to the eigenvalue
$\kappa_m$ given by Eq.\ (\ref{modes}),
\begin{equation}\label{norm}
\int_0^1 dy\,\bar{u}_m(\bar{y})\bar{u}_n(\bar{y}) = \delta_{mn}\,.
\end{equation}
Substitution of Eq.\ (\ref{forced}) into Eq.\ (\ref{eq-bar}) gives
\begin{equation}
\sum_m\left(\frac{d^2R_m}{d\bar{t}^2} + \bar{\omega}_m^2
R_m\right)\bar{u}_m(\bar{y}) = \frac{\epsilon}{2} \frac{d}{d
\bar{t}}\frac{d}{d\bar{y}}[\sigma_{x}(\bar{t})\delta(\bar{y} -
\bar{y}_0)]\,,
\end{equation}
where we have used Eq.\ (\ref{eq-u}). Multiplying both parts of
this equation by $\bar{u}_n(\bar{y})$ and integrating over
$\bar{y}$ from $0$ to $1$ with account of Eq.\ (\ref{norm}), one
obtains linear second-order differential equation for
$R_n(\bar{t})$,
\begin{equation}\label{R-eq}
\frac{d^2R_n}{d\bar{t}^2} + \bar{\omega}_n^2 R_n =
-\frac{\epsilon}{2}\left(\frac{d \sigma_{x}}{d\bar{t}}\right)
\bar{u}'_n(\bar{y}_0)\,,
\end{equation}
where $\bar{u}'_n(\bar{y}_0) \equiv (d\bar{u}_n/d\bar{y})_{\bar{y}
= \bar{y}_0}$.

\begin{figure}[ht]
\begin{center}
\vspace{-0.5cm}
\includegraphics[width=65mm,angle=-90]{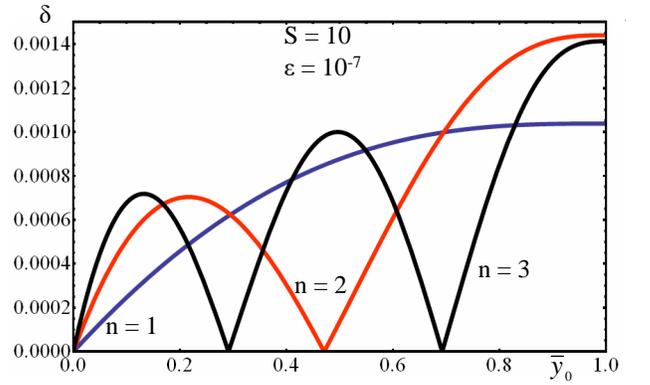}
\vspace{-1cm} \caption{Color online: The dependence of the
relative splitting of the first three cantilever modes on the
position of the molecule.} \label{delta}
\end{center}
\end{figure}

Coupled magneto-mechanical oscillations near the ground state
correspond to small $\sigma_x$, $\sigma_y$, $R_n$, and $\sigma_z
\approx 1$. This requires temperatures $k_BT \ll \Delta$. In this
case Eq.\ (\ref{LL-sigma}) gives for $\sigma_x$
\begin{equation}\label{sx-eq}
\frac{d^2\sigma_x}{d\bar{t}^2} +\bar{\Delta}^2\sigma_x =
S\bar{\Delta}\bar{u}'_n(\bar{y}_0) \frac{dR_n}{d\bar{t}}
\label{sx} \,,
\end{equation}
where $\bar{\Delta} \equiv \Delta/(\hbar \nu)$. Substituting into
Eqs.\ (\ref{R-eq}) and (\ref{sx}) $\sigma_x(\bar{t}), R_n(\bar{t})
\propto \exp(i\bar{\omega}\bar{t})$, one obtains the following
equation for the eigenmodes of the coupled oscillations:
\begin{equation}\label{eigenmodes}
(\bar{\omega}^2 - \bar{\omega}_n^2)(\bar{\omega}^2 -
\bar{\Delta}^2) = \frac{1}{2}\epsilon S
\bar{\Delta}\bar{u}'^2_n(\bar{y}_0)\bar{\omega}^2\,.
\end{equation}
Due to the smallness of $\epsilon$, oscillations of the spin and
the cantilever occur independently at frequencies $\Delta/\hbar$
and $\omega_n$, respectively, unless these two frequencies are
very close to each other. The latter can be achieved by, e.g.,
changing $\Delta$ with the help of the dc magnetic field
perpendicular to the anisotropy axis of the molecule. At $\Delta =
\hbar \omega_n$ one should observe the splitting of the mechanical
mode of the cantilever, $\omega_n$, into two modes
\begin{equation}\label{res-pm}
\omega_{n\pm} = \omega_n\left(1 \pm \frac{\delta}{2}\right)\,,
\quad \delta = \sqrt{\frac{\epsilon S
\bar{u}'^2_n(\bar{y}_0)}{2\bar{\omega}_n}}\,.
\end{equation}
The remarkable property of Eq.\ (\ref{res-pm}) is that it has no
free parameters. For a chosen resonance $\Delta = \hbar \omega_n$,
the relative splitting $\delta$ depends only on the position of
the molecule on the cantilever. This dependence is plotted in
Fig.\ \ref{delta}.

We shall now demonstrate that the above experiment can be
performed by studying the response of the system to a weak ac
magnetic field, ${\bf B}(t) = B_0{\bf e}_x\sin(\omega t)$, applied
along the anisotropy axis of the magnetic molecule. In the
presence of such a field Eq.\ (\ref{b-eff}) becomes
\begin{equation}\label{b-eff-b}
{\bf b}_{\rm eff} = -\Delta {\bf e}_z + 2 S \Delta \delta \phi_x
{\bf e}_y + b{\bf e}_x \sin(\omega t)\,,
\end{equation}
where $b = g\mu_B B_0$, with $g$ being the gyromagnetic factor for
the spin and $\mu_B$ being the Bohr magneton. To obtain the
amplitude of the forced oscillations of the cantilever we need to
include dissipation in the equations of motion of the spin and the
cantilever. The damping modifies the Landau-Lifshitz equation
\cite{book}:
\begin{equation}\label{LL-damping}
\hbar \dot{\bm \sigma} = -{\bm \sigma} \times {\bf b}_{\rm eff} +
2Q_s^{-1}{\bm \sigma} \times \left({\bm \sigma} \times {\bf
b}_{\rm eff}\right)\,.
\end{equation}
Here $Q_s$ is the quality factor of the spin oscillations. Small
oscillations $(|\sigma_{x,y}| \ll 1, \sigma_z \approx 1)$ now
satisfy
\begin{eqnarray}\label{smp-eq}
\frac{d\sigma_{\pm}}{d\bar{t}} + \bar{\Delta}\left(Q_s^{-1}\pm i
\right)\sigma_{\pm} = S\bar{\Delta}\bar{u}'_n(\bar{y}_0) R_n \mp i
\bar{b}\sin(\bar{\omega}\bar{t})
\end{eqnarray}
where $\sigma_{\pm} = \sigma_x \pm i\sigma_y$ and $\bar{b} \equiv
b/(\hbar \nu)$.

Dissipation of the mechanical motion of the cantilever can be
introduced by adding the first time derivative of $R_n$ to Eq.\
(\ref{R-eq}):
\begin{equation}\label{R-dis}
\frac{d^2R_n}{d\bar{t}^2}+
\frac{\bar{\omega}_n}{Q_n}\frac{dR_n}{d\bar{t}} + \bar{\omega}_n^2
R_n = -\frac{\epsilon}{4}
\bar{u}'_n(\bar{y}_0)\frac{d}{dt}(\sigma_+ + \sigma_-) \,.
\end{equation}
Here $Q_n$ is the quality factor of the oscillations of the
cantilever at the eigenfrequency $\bar{\omega}_n$. One can now
obtain the time dependence of $R_n$ and $\sigma_{\pm}$ by solving
together Eq.\ (\ref{smp-eq}) and Eq.\ (\ref{R-dis}). The
displacement of the cantilever at a point $y$ is given by
$\bar{u}_z(\bar{t},\bar{y}) = R_n(\bar{t})\bar{u}_n(\bar{y})$. The
simplest way to get the solution is to replace $i\sin(\bar{\omega}
\bar{t})$ in Eq.\ (\ref{smp-eq}) with $\exp(i\bar{\omega}
\bar{t})$ and solve the resulting three linear algebraic equations
for $R_n, \sigma_{\pm} \propto \exp(i\bar{\omega} \bar{t})$. The
applicability of the formulas obtained that way is limited by the
range of parameters that provides the condition $|\sigma_{\pm}|
\ll 1$ used to derive Eq.\ (\ref{smp-eq}) from Eq.\
(\ref{LL-sigma}). It is easy to see from Eq.\ (\ref{smp-eq}) that
this requirement is violated for $\sigma_-$ when $\omega$ is close
to $\Delta$: The strong pumping of the spin excitations by the ac
field at $\omega \rightarrow \Delta$ leads to the breakdown of the
linear approximation for the dynamics of the spin governed by Eq.
(\ref{LL-sigma}).

At $\omega = \omega_n \neq \Delta$, in the practical range of the
quality factors, $1 \ll Q_{n,s} \ll 1/\epsilon$, one obtains
\begin{equation}\label{u-s}
|\bar{u}_z(\bar{y}_0,\bar{y})| = \frac{\bar{b}\epsilon Q_n
\bar{\Delta}|\bar{u}'_n(\bar{y}_0)|
|\bar{u}_n(\bar{y})|}{2\bar{\omega}_n|\bar{\omega}_n^2
-\bar{\Delta}^2|} \,, \quad |\sigma_{\pm}| =
\frac{\bar{b}}{|\bar{\omega}_n \pm \bar{\Delta}|}\,.
\end{equation}
The parameter $\Delta$ can be controlled by a dc magnetic field
applied perpendicular to the anisotropy axis of the molecule. The
condition $|\sigma_-| \ll 1$ determines how close to the double
resonance, $\Delta = \omega_n$, one can use Eq.\ (\ref{u-s}):
$|\bar{\omega}_n - \bar{\Delta}| \gg \bar{b}/2$. Substitution of
$|\bar{\omega}_n - \bar{\Delta}| \sim \bar{b}$ into the first of
Eqs.\ (\ref{u-s}) provides the estimate for the maximum amplitude
of the oscillations of the cantilever at $\Delta \rightarrow
\omega_n $:
\begin{equation}\label{max}
\max|\bar{u}_z(\bar{y}_0,\bar{y})| \sim \frac{\epsilon
Q_n}{4\bar{\omega}_n}|\bar{u}'_n(\bar{y}_0)|
|\bar{u}_n(\bar{y})|\,.
\end{equation}
It is a function of the distance $y$ from the fixed end of the
cantilever, parameterized by the position of the molecule $y_0$.

For, e.g., $n = 1$ the parameter $\delta$ of Eq. (\ref{res-pm})
should be generally of order $\sqrt{\epsilon S}$. A cantilever of
dimensions 100nm$\times$10nm$\times$1nm, carrying a magnetic
molecule of spin $10$, will have $\nu \sim 10^8$s$^{-1}$,
$\epsilon \sim 10^{-7}$, $\delta \sim 10^{-3}$, see Eqs.\
(\ref{bar}), (\ref{epsilon}), and (\ref{res-pm}). This would give
$f_1 =3.52\nu/(2\pi) \sim 30$MHz and the splitting, $\delta f_1$,
of the first harmonic of the cantilever in the ballpark of a few
kilohertz. The condition for the detection of the splitting is
$Q_{n,s} \gg 1/\delta$. For a single magnetic molecule on a
microcantilever the discrete character of the cantilever phonon
modes and the absence of the inhomogeneous broadening of the spin
mode (usually present in a bulk molecular magnet) should provide a
high spin quality factor. Very high quality factors (up to $10^5$)
have also been reported for microcantilevers \cite{Schwab,Finot}.
When the molecule is near the tip of the cantilever, Eq.\
(\ref{max}) with $\epsilon \sim 10^{-7}$ and $Q_n \sim 10^5$ gives
for the tip: $|\bar{u}_z| \sim 0.005$. For a $100$-nm long
cantilever this would correspond to the oscillations by half a
nanometer. Such a displacement can be detected by tunneling or
force microscopy, as well as by optical and electrical methods.
Working with even smaller cantilevers would provide higher values
of $\epsilon$ and would allow to relax the requirement on the
quality factor.

In Conclusion, we have shown that driven quantum oscillations of
the spin of a magnetic molecule can be observed by placing the
molecule on a microcantilever. Since such cantilevers consist of
hundreds of thousands of atoms, this would be a remarkable example
of a macroscopic quantum effect. Our theory has no free parameters
and, therefore, it must be helpful in designing the proposed
experiment.

The authors thank D. A. Garanin for discussions. This work has
been supported by the NSF Grant No. DMR-0703639.

\end{document}